\documentclass[10pts,showpacs,preprint,aps]{revtex4}
\usepackage{graphicx}
\usepackage{dcolumn}
\usepackage{bm}
\usepackage{amssymb}
\usepackage{amsmath}
\begin{document}
\setcounter{page}{1}
\vskip 2cm
\title
{Vacuum degeneracy in Massive Gravity=Multiplicity of fundamental scales}
\author
{Ivan Arraut$^{(1)}$}
\author 
{Kaddour Chelabi$^{(2)}$}
\affiliation{$^1$ Department of Physics, Faculty of Science, Tokyo University of Science,
1-3, Kagurazaka, Shinjuku-ku, Tokyo 162-8601, Japan}
\affiliation{$^2$ State Key Laboratory of Theoretical Physics, Institute of Theoretical Physics, Chinese Academy of Science, Beijing 100190, Peoples Republic of China}

\begin{abstract}
The presence of Nambu-Goldstone bosons introduce a natural degeneracy inside the vacuum solutions of the non-linear formulations of massive gravity in the same spirit of the $\sigma$-models. When the gravitational effects are taken into account, and the observers are located at any distance with respect to the source, this degeneracy corresponds to a multiplicity (flow) of the fundamental scales of the theory. The different values of the fundamental scales are connected each other through the broken generators of the theory.  
\end{abstract}
\maketitle 
\section{Introduction}
In order to solve the problem of dark energy, different modified gravity theories have been proposed. All of them have something in common, namely, the appearance of an additional scale and the introduction of non-derivative terms interactions in the action. Among the most popular approaches in order to modify gravity, we have massive gravity theories \cite{Mass2, deRham, deRham2}, Hordenski theories \cite{Horde} and others. The reality is that although modified gravity theories show a common pattern of behavior, they are not fundamental in the sense that they will not explain in a deep sense the origin of the accelerated expansion of the universe. The importance of modified gravity theories is that they can provide observables that can be confronted with experiments. What distinguish massive gravity theories from other approaches is the breaking of the diffeomorphism invariance for the vacuum solutions, such that the theory contains five degrees of freedom instead of the traditional two degrees of freedom which propagate General Relativity (GR). The diffeomorphism invariance is broken explicitly by introducing a second non-dynamical metric dubbed "fiducial". The diffeomorphism invariance of the action is restored by introducing St\"uckelberg fields. There are two ways of using the St\"uckelberg trick. The first one is by introducing the extra-degrees of freedom inside the fiducial metric. In such a case, the dynamical metric can take the form of the standard solutions of GR. This is the way analyzed in most of the massive gravity papers \cite{Mass2, deRham, deRham2}. The second method, is to introduce the St\"uckelberg fields inside the dynamical metric and then keeping the fiducial metric fixed as Minkowski. In this second method, it is impossible for the dynamical metric to agree with the standard results of GR. This is the method formulated in \cite{K} and used widely in \cite{Vainshtein paper, HawkingdRGT, HawkingdRGT2, Kodama, Higgs, Higgs2, My paper1, My paper12, My paper1la, Komar, V1, V2, V3, V4}. The introduction of the extra-degrees of freedom inside the dynamical metric facilitates the Hawking radiation analysis \cite{Vainshtein paper, HawkingdRGT, HawkingdRGT2} for the black hole solutions \cite{Kodama}. In addition, it provides an excellent scenario for a consistent formulation of a graviton Higgs mechanism \cite{Higgs, Higgs2}. In this paper, we analyze the vacuum solutions for the spherically symmetric case in an arbitrary frame of reference. In this sense, the gravitational effects of the source will be introduced perturbatively through the dimensionless parameter $\epsilon$, which depends on the fundamental scales of the theory and on the distance between the observer and the source. We demonstrate that the vacuum degeneracy of the theory makes it impossible to fix the fundamental scales of the theory. If we imagine a set of observers located at the same position in space but defining different notions of time $t$, then they will define different fundamental scales for the same theory. The observers defining different notions of time are connected through the broken generators of the theory. As a consequence of this, the flow of the fundamental scales of the theory is governed by the broken generators. For the special case where only the scalar component of the St\"uckelberg function is taken into account, the broken generators are of the form $U(1)$. In such a case, the $U(1)$ symmetry transformations then connect the different values of the fundamental scales of the theory.          

\section{Massive gravity theories}

Massive gravity theories can be formulated by the generic action \cite{Mass2, deRham, deRham2, Kodama}

\begin{equation}   \label{eq:b1}
S=\frac{1}{2\kappa^2}\int d^4x\sqrt{-g}(R+m_g^2U(g,\phi)),
\end{equation}
where the first term is just the standard Einstein-Hilbert action and the second term corresponds to the partial contractions between the dynamical and fiducial metric. Here the potential $U(g,\phi)$ is defined as

\begin{equation}   \label{eq:b2}
U(g,\phi)=U_2+\alpha_3 U_3+\alpha_4U_4.
\end{equation}
Note that the potential has two free-parameters in addition to the graviton mass parameter represented by $m_g$. The explicit form of the potential is given by

\begin{equation}   \label{eq:b3}
U_2=Q^2-Q_2,
\end{equation}

\begin{equation}   \label{eq:b4}
U_3=Q^3-3QQ_2+2Q_3,
\end{equation}

\begin{equation}   \label{eq:b5}
U_4=Q^4-6Q^2Q_2+8QQ_3+3Q_2^2-6Q_4,
\end{equation}

\begin{equation}   \label{eq:b6}
Q=Q_1,\;\;\;\;\;Q_n=Tr(Q^n)^\mu_{\;\;\nu},
\end{equation}

\begin{equation}   \label{eq:b7}
Q^\mu_{\;\;\nu}=\delta^\mu_{\;\;\nu}-M^\mu_{\;\;\nu},
\end{equation}

\begin{equation}   \label{eq:b8}
(M^2)^\mu_{\;\;\nu}=g^{\mu\alpha}f_{\alpha\nu},
\end{equation}

\begin{equation}   \label{eq:b9}
f_{\mu \nu}=\eta_{ab}\partial_\mu\phi^a\partial_\nu\phi^b.
\end{equation}
Then the field equations are given by 

\begin{equation}   \label{eq:b10}
G_{\mu \nu}=-m^2X_{\mu\nu},
\end{equation}
where

\begin{equation}   \label{eq:b11}
X_{\mu \nu}=\frac{\delta U}{\delta g^{\mu \nu}}-\frac{1}{2}Ug_{\mu \nu}.
\end{equation}
Here $f_{\mu \nu}$ is the fiducial metric and $Q$ is the trace of the matrix $Q^\mu_{\;\;\nu}$ taken with respect to the dynamical metric. 

\section{Spherically symmetric solutions}

The solutions for the previous equations in the spherically symmetric case, can be expressed in a generic form as

\begin{equation}   \label{eq:drgtmetric222miau}
ds^2=-f(S_0r)dT_0(r,t)^2+\frac{S_0^2}{f(S_0r)}dr^2+S_0^2r^2d\Omega_2^2,
\end{equation}
with

\begin{equation}    \label{eq:drgtmetric222miau2}
dT_0(r,t)=\dot{T}_0(r,t)dt+T_0'(r,t)dr.
\end{equation}
Note that $T_0(r,t)$ corresponds to the St\"uckelberg function, which contains the fundamental scales of the theory and the information of the extra-degrees of freedom of the theory. The St\"uckelberg function appears when we apply the St\"uckelberg trick on the dynamical metric in the form 

\begin{equation}   \label{eq:stuefashio}
g_{\mu\nu}=\left(\frac{\partial Y^\alpha}{\partial x^\mu}\right)\left(\frac{\partial Y^\beta}{\partial x^\nu}\right)g'_{\alpha\beta},
\end{equation}  
where in general 

\begin{equation}   \label{eq:stuefashio22}
Y^\alpha(x)=x^\alpha+A^\alpha(x),
\end{equation}
with the components of the St\"uckelberg function given by

\begin{equation}   \label{eq:stuefashio2}
Y^0(r,t)=T_0(r,t),\;\;\;\;\;Y^r(r,t)=S_0r.
\end{equation}
Then 

\begin{equation}   \label{eq:tsubzerla}
T_0(r,t)= S_0t+A^t(r,t),\;\;\;\;\;\;\; A^r(r,t)=0. 
\end{equation}
Note that the radial component is trivial and it does not contain any information about the extra-degrees of freedom. For the time-component, represented by $T_0(r,t)$, the situation is different since it is a non-trivial function. The gravitational effects will appear through the function $f(S_0r)$ defined as follows \cite{Kodama, Higgs, Higgs2}

\begin{equation}   \label{eq:observerlocated}
f(S_0r)=1-\frac{2GM}{S_0r}-\frac{1}{3}\Lambda S_0^2r^2.
\end{equation}
Note that $S_0$ is a function of the two free-parameters of the theory and the cosmological constant ($\Lambda$) also depends on the two free-parameters and on the graviton mass parameter $m_g$ \cite{Kodama}. The gravitational effects can be interpreted as a deviation from the Minkowski space and they will appear from deviations with respect to one for the function $f(S_0r)$ in eq. (\ref{eq:drgtmetric222miau}) and from the non-trivial deviations of the function $T_0(r,t)$ with respect to the usual notion of time $t$. This last statement can be perceived from eq. (\ref{eq:tsubzerla}) together with the solution (\ref{eq:drgtmetric222miau}) and the result (\ref{eq:drgtmetric222miau2}). We can then define a  dimensionless parameter $\epsilon$ which depends on the fundamental scales of the theory and on the location of the observer with respect to the source as follows 

\begin{equation}   \label{eq:epsilon def.}    
f(S_0r)=1-\epsilon.
\end{equation}
The exact form of $\epsilon$ is not important at this point. What is really relevant is the fact that it is a parameter containing the fundamental scales of the theory, namely, $G$ and $m_g$. The parameter is defined once we fix the location of the observer with respect to the source. Then for practical purposes, we will assume that $\epsilon$ only depends on the fundamental scales of the theory.   

\section{Vacuum solutions in the absence of gravity}   \label{Vacsol}

Massive gravity is in essence a sigma model, then the vacuum solution of the theory will be single or degenerate depending on the combination of the two free-parameters of the theory \cite{Publishedone}. The two free-parameters defined in eq. (\ref{eq:b2}), can be re-expressed in terms of a new set of parameters $\alpha$ and $\beta$, here defined as \cite{Kodama}

\begin{equation}
\alpha=1+3\alpha_3,\;\;\;\beta=3(\alpha_3+4\alpha_4).
\end{equation}
In \cite{Publishedone}, the two family of black-hole solutions were classified depending on the relation between the two free-parameters of the theory $\alpha$ and $\beta$. Then the analogy with respect to the non-linear sigma models was studied. It was then concluded that the solutions Type II, belonging to the special combination $\beta=\alpha^2$, correspond to the solutions where the vacuum is degenerate. This degeneracy comes from the arbitrariness of the function $T_0(r,t)$. Interpreted as a preferred time-direction, $T_0(r,t)$ defines the preferred notion of vacuum, where the graviton mass effects are absent. Observers defining different notions of time $t$, will define different notions of vacuum where the graviton mass effects can be perceived. The non-trivial derivatives of the St\"uckelberg function, obtained from eq. (\ref{eq:tsubzerla}) are defined by  

\begin{equation}   \label{eq:tprime}
T_0'(r,t)=A'(r,t),\;\;\;\dot{T}_0(r,t)=S_0+\dot{A}^t(r,t).
\end{equation}
Here $T_0'(r,t)$ represents the spatial derivative of the St\"uckelberg function.

\subsection{The case of one free-parameter $\beta=\alpha^2$: Vacuum degenerate}

Here we will only focus for the case of one free-parameter where it has been demonstrate before that the vacuum solution is degenerate \cite{Kodama, Publishedone}. Note that for a stationary background, the time derivative of the function $A^t(r,t)$ only appears at the perturbative level, then we can conclude that $\dot{A}^t(r,t)<<S_0$. The root square of the determinant of the dynamical metric is expanded as

\begin{equation}   \label{eq:deltaU1}
\sqrt{-g}\approx S_0^3\left( S_0+\dot{A}^t(r,t)\right)\left(1+\frac{1}{2}h-\frac{1}{4}h^\alpha_{\;\;\beta}h^\beta_{\;\;\alpha}+\frac{1}{8}h^2\right).
\end{equation}
The perturbation of the other part of the potential defined by $U(g, \phi)$, is given by

\begin{eqnarray}   \label{eq:deltaU} 
\delta U(g,\phi)=\frac{(1+\alpha)^2}{S_0^3\alpha^3}h_{00}(r,t)-\frac{3(1+\alpha)^2}{S_0^4\alpha^3}\dot{A}^t(r,t)h_{00}(r,t)
-\frac{2(1+\alpha)^2}{S_0^2\alpha^2}\dot{A}^t(r,t)\nonumber\\
+(\dot{A}^{t}(r,t))^2F_1(\alpha,S_0)
-F_2(\alpha,S_0)T_0'(r,t)h_{0r}(r,t)
-\frac{S_0(1+\alpha)^4}{S_0^2\alpha^5}h_{rr}(r,t)\nonumber\\
+\frac{(1+\alpha)^4}{S_0^2\alpha^5}\dot{A}^t(r,t)h_{rr}(r,t).
\end{eqnarray} 
The background potential is defined by

\begin{equation}   \label{eq:deltaU2}
U(g,\phi)_{back}= -\frac{2}{\alpha}+\frac{2(1+\alpha)^2}{S_0\alpha^2},
\end{equation}
with $\dot{T}_0(r,t)=S_0+\dot{A}^t(r,t)$ and $\dot{A}^t(r,t)$ representing the perturbative deviations with respect to the stationary condition. Joining the results (\ref{eq:deltaU1}), (\ref{eq:deltaU}) and (\ref{eq:deltaU2}), we then obtain the expression for the full potential $V(g, \phi)$, defined as

\begin{equation}   \label{eq:b11pot}
V(g,\phi)=\sqrt{-g}U(g,\phi)),
\end{equation}
and explicitly given by

\begin{eqnarray}
V(g,\phi)\approx S_0^4\left(1+\frac{1}{2}h+\frac{1}{4}h_{\mu\nu}^2+\frac{1}{8}h^2\right)U(g,\phi)_{back}
+S_0^4\left(1+\frac{1}{2}h\right)\delta U(g,\phi)\nonumber\\+S_0^3\dot{A}^t\left(1+\frac{1}{2}h\right)U(g,\phi)_{back}+S_0^3\dot{A}^t\delta U(g,\phi).
\end{eqnarray} 
The vacuum for the perturbations of the metric is defined by

\begin{equation}   \label{eq:acalanojo}
\frac{dV(g,\phi)}{dh_{\mu \nu}}=0.
\end{equation}
and then we have a general solution depending on the parameter $\alpha$ and the functions $\dot{T}_0(r,t)$ and $T_0'(r,t)$ as follows

\begin{equation}   \label{eq:thisvacuum too}
h_{\mu\nu vac}=F_{\mu\nu}(\alpha, \dot{A}^t(r,t), T_0'(r,t)).
\end{equation}
Here $F_{\mu\nu}$ is just a tensorial function summarizing the explicit results. The exact result for $h_{\mu\nu vac}$ is not relevant at this point. What is important is to understand that the vacuum is degenerate because the function $T_0(r,t)$ is degenerate for the solution under consideration. In \cite{Kodama, Higgs, Higgs2}, this degeneracy was reported and it was the key point for the formulation of the graviton Higgs mechanism. Note that if $T_0(r,t)$ is uniquely defined as it is the case for the {\it Type I} solutions defined in \cite{Publishedone}, then the vacuum defined by eq. (\ref{eq:thisvacuum too}) is unique after fixing the parameters of the theory.          

\section{Vacuum solutions in the presence of gravity}   \label{Vacsol2}

When the gravitational effects are included, it is impossible to fix the fundamental scales of gravity, namely, $G$ or $m$ for a degenerate vacuum. Here again we define the potential for gravity as it was defined in eq. (\ref{eq:b11pot}). For this case, the result (\ref{eq:tprime}) is still valid. However, the spatial derivative of the St\"uckelberg function will contain now the fundamental scales of the theory as has been demonstrated in \cite{Publishedone}. Since massive gravity is a sigma model, it is known that if $\beta\neq\alpha^2$, then the vacuum solution will be uniquely defined. It has been demonstrated in \cite{Kodama} that the for this case, the St\"uckelberg function will be constrained to obey some specific behavior, defined by 

\begin{equation}   \label{eq:wonderful}
(T_0'(r,t))^2=\frac{S_0^2(1-f(S_0r))}{f(S_0r)}\left(\frac{1}{f(S_0r)}-1\right),
\end{equation}
for the cases where the solution obeys the spherical symmetry. The solutions satisfying this constraint were dubbed as $Type\; I$ in \cite{Publishedone}. If we develop the series expansion as it was done in \cite{Publishedone}, and keeping then the series at the lowest order, then we obtain the result

\begin{equation}   \label{eq:thislamiau}
\vert T_0'(r,t)\vert \approx S_0 \epsilon,
\end{equation}
for $\epsilon<<1$, which is the usual case for observers located at scales far away from any horizon (event horizon or cosmological horizon). Note that since $\epsilon$ contains the fundamental scales of the theory, then for the case of solutions $Type\; I$, the fundamental scales of the theory will be fixed and they will not flow. By continuity in the flux of parameters between the solutions $Type\; I$ and $Type\; II$, then it is valid to take the result (\ref{eq:thislamiau}) as the appropriate one for the solutions $Type\; II$, but taking into account that this time $\epsilon$ must be arbitrary. In such a case, then the fundamental scales of the theory are never fixed because the vacuum for this case is degenerate. This is the interesting situation for the purposes of the paper. The potential (\ref{eq:b11pot}) expanded up to second order in perturbations is given by a function with the following dependence

\begin{equation}   \label{Massive action}
V(g,\phi)= V(\alpha, h_{\mu\nu}, \epsilon, \dot{A}^t(r,t)),
\end{equation}
where $\alpha$ is the free-parameter of the theory, $\epsilon$ is the dimensionless scale of gravity which enters perturbatively, $\dot{A}^t(r,t)$ corresponds to the non-trivial time derivative of the St\"uckelberg function $T_0(r,t)$, defined by the results (\ref{eq:tprime}). $h_{\mu\nu}$ represents the graviton field. Note that by eq. (\ref{eq:tprime}), $\dot{A}^t(r,t)$ represents the deviations with respect to the stationary condition at the perturbative level. As has been just mentioned, here the degeneracy of the vacuum is translated to the scale $\epsilon$, which depends on the Newtonian constant $G$ and on the graviton mass $m_g$.  

\section{Broken symmetry and degeneracy of the fundamental scales}

If we solve the equation (\ref{eq:acalanojo}) for the graviton perturbations $h_{\mu\nu}$, then we obtain a function of the form

\begin{equation}   \label{eq:hmunu}
h_{\mu\nu vac}=F_{\mu\nu}(\alpha, \dot{A^t}(r,t), \epsilon),
\end{equation}
where $F_{\mu\nu}(x)$ is a function of the argument $x=\alpha, \dot{A}^t(r,t), \epsilon$. Here we can write the explicit result for the $0-0$ component as follows

\begin{eqnarray}  
h_{00 vac}=C_0(\alpha)+C_1(\alpha)\dot{A}^t(r,t)+C_2(\alpha)\dot{A}^t(r,t)^2\nonumber\\
+C_3(\alpha)\epsilon+C_4(\alpha)\epsilon^2+C_5(\alpha)\epsilon\dot{A}^t(r,t)+....
\end{eqnarray}
Here $C_n$ are just constants. Analogous results can be obtained for the other components. Note the expansion in terms of the function $\epsilon$. The vacuum obtained in this way, is naturally degenerate in the sense that it is not invariant under the set of diffeomorphism transformations related to the St\"uckelberg fields and defined as follows 

\begin{equation}   \label{eq:nontrivial}
\delta_g T_0(r,t)=-\zeta(Y) \approx-\zeta^t-A^\alpha\partial_\alpha\zeta^t-\frac{1}{2}A^\alpha A^\beta\partial_\alpha\partial_\beta\zeta^t+...,
\end{equation}
where $\zeta^t=\delta_g t$ and $A^\alpha$ is defined by eq. (\ref{eq:tsubzerla}). Once we fix the location of the observer with respect to the source, then $\epsilon$ only depends on the fundamental scales of the theory. If we define the St\"uckelberg function by the result (\ref{eq:thislamiau}), then its arbitrariness is equivalent to an arbitrariness of $\epsilon$ and as a consequence into an arbitrariness of the fundamental scales of the theory. This type of degeneracy is then translated to the vacuum solutions defined by eq. (\ref{eq:hmunu}). An interesting fact about the transformations (\ref{eq:nontrivial}) is that those corresponding to the trivial case where the St\"uckelberg fields $A^\alpha(x)$ vanishes, in other words, those transformations defined trivially by $\zeta^t=\delta_g t$ (as in standard GR), will connect equivalent vacuums. For these special type of transformations, the vacuum will be uniquely defined. On the other hand, the vacuum defined by eq. (\ref{eq:hmunu}) is not invariant under the full set of transformations (\ref{eq:nontrivial}) including the St\"uckelberg fields. We can imagine that different observers can define different set of vacuums connected each other through the set of transformations (\ref{eq:nontrivial}) and that each observer then define different values of the fundamental scales of the theory. In fact, the full set of transformations (\ref{eq:nontrivial}) correspond to the set of broken generators of the theory and they are in principle related to the number of Nambu-Goldstone bosons. By the date however, it is not clear what is the exact connection between the number of Nambu-Goldstone bosons and the number of broken generators when we are talking about spacetime symmetries. We can extend the St\"uckelberg field's definition by the introduction of the $U(1)$ symmetry as follows \cite{K}

\begin{equation}   \label{eq:transformationla}
A_\alpha\to A_\alpha+\partial_\alpha\phi.
\end{equation}
In such a case, the transformation defined in eq. (\ref{eq:nontrivial}), is extended and it becomes

\begin{eqnarray}   \label{eq:nontrivial2}
\delta_g T_0(r,t)=-\zeta(Y) \approx\partial^t\lambda(x)-\zeta^t-A^\alpha\partial_\alpha\zeta^t-\frac{1}{2}A^\alpha A^\beta\partial_\alpha\partial_\beta\zeta^t+...,\nonumber\\
\delta_g\phi=-\lambda(x).
\end{eqnarray}
For the special case where the St\"uckelberg fields are reduced to the scalar case $\phi$, then the previous transformations are

\begin{eqnarray}   \label{eq:nontrivial21}
\delta_g T_0(r,t)=-\zeta(Y) \approx \partial^t\lambda(x)-\zeta^t,\nonumber\\
\delta_g\phi=-\lambda(x).
\end{eqnarray}
For this special case, then the definition of the time by the observers is reduced to the way how the scalar field $\phi$ is defined. The field $\phi$ in a complex plane transform in agreement with the the group $U(1)$. Then for this special case, the fundamental scales of the theory are connected through the $U(1)$ symmetry transformations of the broken generators. $U(1)$ corresponds then to the set of broken generators of the theory.  
 
\section{The arbitrariness of the fundamental scales of the theory}        

\begin{figure}
	\centering
		\includegraphics[width=15cm, height=8cm]{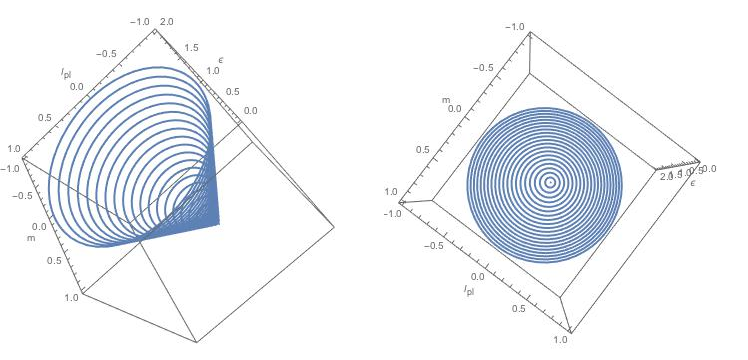}
	\caption{The flow of the fundamental scales of the theory as a function of $\epsilon$. The flow describes a conical behavior. The figure on the right is the flow observed from the $l_{pl}-m_g$ plane.}
	\label{fig1K}
\end{figure}

In order to illustrate why the fundamental scales of the theory are arbitrary when the vacuum is degenerate, from eqns. (\ref{eq:observerlocated}) and (\ref{eq:epsilon def.}), we can express the parameter $\epsilon$ as

\begin{equation}   \label{eq:epsilonlala}    
\epsilon=\frac{2GM}{S_{eff}}+\frac{1}{3}\Lambda S^2_{eff}.
\end{equation}
Here $S_{eff}=S_0r$ and $\Lambda$ is a function of the graviton mass and of the free-parameter $\alpha$ as has been defined in \cite{Kodama}. For solutions $Type \;II$, it has been demonstrated in \cite{Kodama} that $\Lambda=m^2/\alpha$. If $T_0(r,t)$ is arbitrary as it is the case for $Type \;II$ solutions, then by using eq. (\ref{eq:thislamiau}), $\epsilon$ is arbitrary and then the combination given in eq. (\ref{eq:epsilonlala}) is arbitrary. If we fix the mass of the source $M$, and the parameter $\alpha$, then the dependence of the function (\ref{eq:epsilonlala}), becomes 

\begin{equation}   \label{eq:epsiloreal}
\epsilon=A(\alpha)l_{Pl}^2+B(\alpha)m_g^2,
\end{equation}
where we have defined $G=l_{pl}^2$ in connection with the Newtonian constant and here $A(\alpha)$ and $B(\alpha)$ are functions depending on the parameter $\alpha$. If we apply eq. (\ref{eq:thislamiau}), then from eq. (\ref{eq:epsiloreal}), we conclude that

\begin{equation}   \label{eq:thislamiau55}
\vert T_0'(r,t)\vert \approx C(\alpha)l_{Pl}^2+D(\alpha)m_g^2.
\end{equation}	
where again $C(\alpha)$ and $D(\alpha)$ are functions depending on the free parameter $\alpha$. The flow of the fundamental scales of the theory can be observed from Fig. (\ref{fig1K}). Note that the arbitrariness of the fundamental scales affect the way how we define some astrophysical scales of the theory. For example, the Vainshtein radius in massive gravity, defined as $r_v\backsim (GM/m_g^2)^{1/3}$ would become arbitrary under the situation where the vacuum is degenerate. This means that different observers would define different Vainshtein scales depending on how they define their time-direction with respect to $T_0(r,t)$. Take into account that the Vainshtein scale is analogous to the scale $r_0=(3GMr_\Lambda^2)^{1/3}$ ($r_\Lambda=1/\sqrt{\Lambda}$), obtained inside the scenario of ordinary gravity with a non-zero cosmological constant \cite{My paper1, My paper12}. Then the vacuum degeneracy is expected to affect the way how we define bound orbits in massive gravity theories.      
 
\section{Massive gravity with Lorentz violation}

We must remark that there are different versions of massive gravity. All of them sharing similar properties. The differences between one or another theory come from the way how the non-derivative terms of the action appears. In other words, what differentiates one theory from the other is the way how we write the massive action. One interesting suggestion appears when the theory breaks explicitly or spontaneously the Lorentz symmetry as has been explained in \cite{Rubakov, Rubakov2}. Here we will compare these two cases with the results obtained in this paper. In order to make the comparison it is enough to analyze the behavior of the theory when the background metric is close to Minkowski. In order to make a comparison with the results obtained in this paper, then it is enough to work in a free-falling frame. This however does not switch-off completely the gravitational effects which will appear from the deviations of time with respect to the preferred time-direction defined by $T_0(r,t)$. In what follows we will separate the analysis in three parts, the first part covers a short summary of theories breaking the Lorentz symmetry explicitly. The second part covers theories breaking the Lorentz symmetry spontaneously. This second part is the most interesting one for the purposes of this paper. Finally, we make some small comments about the ghost condensation models. 

\subsubsection{Lorentz symmetry broken explicitly} 

In this case, the massive action does not respect the symmetry. With the background metric taken as Minkowski, the massive action expanded up to second order in perturbations is

\begin{equation}    \label{R1}
S_m=\frac{M_{pl}^2}{2}\left(m_0^2h_{00}h_{00}+2m_1^2h_{0i}h_{0i}-m_2^2h_{ij}h_{ij}+m_3^2h_{ii}h_{jj}-2m_4^2h_{00}h_{ii}\right).
\end{equation}  
From the massive action (\ref{R1}), it is possible to recover the Fierz-Pauli theory in the limit where the following condition is satisfied

\begin{equation}
m_0^2=0,\;\;\;m_1^2=m_2^2=m_3^2=m_4^2=m^2.
\end{equation}   
It can be demonstrated that the relations between the different parameters of the massive action are connected with the type of symmetries satisfied by it \cite{Rubakov, Rubakov2}. This is an important point because in this paper we have demonstrated that in massive gravity, the relation between the different parameters of the theory define the type of symmetries which are valid. The difference is that in the theories considered here, the symmetry is not broken explicitly but rather spontaneously as has been demonstrated previously. By combining the massive action (\ref{R1}) with the Einstein-Hilbert action, it is possible to obtain the field equations. In \cite{Rubakov, Rubakov2}, the field equations were found for the different modes, namely, tensor mode, vector mode and scalar mode. The ghost mode is absent under some special conditions imposed over the parameters of the massive action (\ref{R1}) as has been explained in \cite{Rubakov, Rubakov2}. Although the theory described in this paper breaks the Lorentz symmetry spontaneously rather than explicitly, it is still possible to obtain the mass parameters for each mode. In a free-falling frame for example, and assuming the ideal stationary condition $\dot{T}_0(r,t)=S_0$ ($\dot{A}^t(r,t)=0$), the massive action (\ref{Massive action}) becomes

\begin{eqnarray}   \label{eq:acalanojo2miau}
V(g, \phi)=\sqrt{-g}U(g,\phi)\approx \left(1+\frac{1}{2}h-\frac{1}{4}h^\alpha_{\;\;\beta}h^\beta_{\;\;\alpha}+\frac{1}{8}h^2\right)\left(\frac{2+6\alpha(1+\alpha)}{(1+\alpha)^4}\right)\nonumber\\
-\left(1+\frac{1}{2}h\right)\left(-\frac{h}{1+\alpha}+\frac{2T_0'(r,t)(1+\alpha)^2}{\alpha^3}h_{0r}-\frac{T_0'(r,t)^2(1+\alpha)^3}{\alpha^4}\right)+...     
\end{eqnarray}
From this result it is possible to evaluate the masses for each mode through the second derivatives of the potential

\begin{equation}   \label{masses}
\left(\frac{\partial^2 V(g, \phi)}{\partial h_{\mu\nu}\partial h_{\alpha\beta}}\right)_{vac}.
\end{equation}        
This result represents the mass matrix, which after diagonalization can give us the appropriate eigenvalues corresponding to the different modes. Note that the matrix is evaluated at the vacuum value. We can perceive without doing explicitly the calculations that the mass for each mode will depend explicitly on $T_0'(r,t)$ which is arbitrary for the case considered here. Since $T_0(r,t)$ represents a preferred notion of time, then an observer defining the time in this direction, will not perceive the effects of $T_0'(r,t)$ in the masses corresponding to the different modes. On the other hand, an observer defining the time arbitrarily, will perceive the effects of $T_0'(r,t)$ and then will define the masses for each mode depending dynamically on the St\"uckelberg functions. Note that the gravitational effects for a free-falling observer come from the deviations between the time defined by the observer and the preferred time-direction defined by the function $T_0(r,t)$.  

\subsubsection{Lorentz symmetry broken spontaneously}

This case corresponds to the situation where the action respects the Lorentz symmetry and at the same time, the symmetry is not satisfied at the background level due to the spacetime dependence of additional scalar fields corresponding to the Goldstone modes \cite{Rubakov, Rubakov2}. This situation has a direct relation with the one showed in this paper because the St\"uckelberg fields take the role of Goldstone modes. As has been explained previously, in order to make a comparison, it is enough to consider the expansions around the free-falling observers. Note however that still the fundamental scales of the theory can appear due to the deviations between the time defined by the observers and the preferred notion of time defined by $T_0(r,t)$. In the analysis showed in \cite{Rubakov, Rubakov2}, the following scalar (Goldstone) fields

\begin{equation}
\phi^0=a\lambda^2t,\;\;\;\;\;\;\;\phi^i=b\lambda^2x^i,
\end{equation}     
are defined. The action for this case respects the Galilean symmetry defined as a shift of the scalars in the same way as it is defined in \cite{Horde, Rubakov, Rubakov2}. Then the massive action in this case, contains only derivatives of the fields $\phi^a$. Note that this case is similar to the one analyzed in this paper where the massive action also contains the derivatives of the fields $\phi^a$. In both cases, namely, in the model analyzed in \cite{Rubakov, Rubakov2}, as well as the one discussed here, the action can be expressed as

\begin{equation}   \label{Ruba2}
S=S_{EH}+S_\phi,
\end{equation} 
where $S_{EH}$ is the Einstein-Hilbert action and $S_\phi$ is the massive action. In \cite{Rubakov, Rubakov2}, the massive action takes the form

\begin{equation}
S_\phi=\int\sqrt{-g}\lambda^4F(X, V^i, Y^{ij}, Q),
\end{equation} 
where $X$, $V^i$, $Y^{ij}$ and $Q$ are defined as a function of the derivatives of $\phi^a$. In this paper, the massive action is defined by eqns. (\ref{eq:b1}) and (\ref{eq:b2}). By expanding such action up to second order in a free-falling frame for the case of one free-parameter (St\"uckelberg function arbitrary), we would obtain again the result (\ref{eq:acalanojo2miau}). We then observe once again that the masses for the modes are related to the derivative of the St\"uckelberg function $T_0'(r,t)$ if we evaluate the result (\ref{masses}). Such masses are again consistent with the definitions of the massive gravity theories violating Lorentz symmetry spontaneously. In \cite{Rubakov, Rubakov2}, it is explained some possible pathologies for the actions of the form (\ref{Ruba2}) and the possibilities for solving them.      
    
\subsubsection{Ghost condensation}

For the models of ghost condensation introduced in \cite{Nima}, the graviton is massless \cite{Rubakov, Rubakov2}. This model is UV-complete and still the action defined generically in the form (\ref{Ruba2}) is valid with 

\begin{equation}   \label{gm}
S_\phi=\lambda^4\int d^4x\sqrt{-g}F(X).
\end{equation}   
Here $X=\lambda^{-4}g^{\mu\nu}\partial_\mu\phi\partial_\nu\phi$ and the solution for $\phi$ is \cite{Rubakov, Rubakov2, Nima}

\begin{equation}   \label{thisoneice}
\phi=\gamma\lambda^2t.
\end{equation}
This term violates time translations and as a consequence the energy is conserved but not in the usual sense. In fact, it is a combination between time-translations plus shifts of the field $\phi$ what defines the conserved energy. This is analogous to the situation explored in \cite{Komar}, where a consistent explanation for the conservation of energy was developed inside the scenario of massive gravity theories. In such a case, it was demonstrated that the energy is conserved even if the symmetry under time-translations was lost. The solution (\ref{thisoneice}) also violates Lorentz symmetry. Further discussion about the ghost condensation models can be found in \cite{Rubakov, Rubakov2, Nima}. In the theory presented in this paper, the notion of preferred time-direction defined by the $T_0(r,t)$, plays an analogous role to the result (\ref{thisoneice}) if we include perturbations in the analysis as has been done in \cite{Rubakov, Rubakov2}. However, it is evident that the ghost-condensation model is in essence different to the theory worked in this paper.       

\section{Conclusions}

In this paper we have showed that in massive gravity, when the vacuum solution is degenerate, it is impossible to fix the fundamental scales of the theory. The vacuum is degenerate for the solutions obeying $\alpha=\beta^2$ which corresponds to the $Type\; II$ solutions. In this paper the gravitational effects are introduced perturbatively through the dimensionless parameter $\epsilon$. The parameter contains the fundamental scales of the theory. If a free-falling observer defines the time in agreement with the St\"uckelberg function $T_0(r,t)$, then he/she will describe a vanishing parameter $\epsilon$ and then the gravitational effects will be absent. Any other type of observer defining the time with a function $t\neq T_0(r,t)$, will define $\epsilon\neq0$ and then the gravitational effects will appear explicitly. The observers defining different notions of time are connected through the set of broken generators of the theory. Then the set of transformations corresponding to the broken generators, can be considered as a flow of the fundamental constants of the theory. This is analogous to the spirit of the Wilson approach to the renormalization where the fundamental scales of the theory also flow. The case presented here is however more complex because it relates the fundamental scales of the theory with the observer conditions and all the observers are connected through the set of broken generators. Then this situation is more interesting than the ones corresponding to the case of internal symmetries. We must also remark that the observers defining different scales define different notions of vacuum. Then an equivalent conclusion from this paper is that the set of broken generators connect different set of vacuums, defining different fundamental scales. Note that the degeneracy of the vacuum appears for the {\it Type\;II} solutions, which corresponds to the case $\beta=\alpha^2$. Note that for $\beta\neq\alpha^2$ the vacuum is unique and it is possible to fix the fundamental scales of the theory. These situations correspond to the description of massive gravity as a gravitational $\sigma$-model as has been defined in \cite{Publishedone}. The theory analyzed in this paper is analogous to the Lorentz violated theory but for the case when the symmetry is spontaneously broken. Similar results have been analyzed in \cite{Rubakov, Rubakov2}. Finally we can conclude that if the fundamental scales of the theory are arbitrary for some combination of parameters, then their combinations generating astrophysical scales will be also arbitrary. As a consequence of this, the notion of bound orbits becomes ambig\"uous at this level. \\\\  

{\bf Acknowledgement}
I. A is supported by the JSPS Post-doctoral fellowship for oversea Researchers. K. C. is supported by the CAS Twast Presidential program for Ph.D. students.


\begin{thebibliography}{0}
\bibitem{Mass2} Fierz M. and Pauli W., Proc.Roy.Soc.Lond. {\bf A173}: 211, (1939).
\bibitem{deRham} de Rham C., Gabadadze G. and Tolley A. J., Phys. Rev. Lett. {\bf 106}, 231101, (2011). 
\bibitem{deRham2}  C. de Rham and G. Gabadadze, Phys.Rev. D{\bf82}, 044020, (2010).
\bibitem{Horde} Horndeski G. W., Int. J. Theor. Phys. {\bf10}, 363-384 (1974).
\bibitem{K} Hinterbichler K., Phys. Rev. D {\bf 84}, 671 (2012). 
\bibitem{Vainshtein paper} I. Arraut, arXiv:1407.7796 [gr-qc].
\bibitem{HawkingdRGT} I. Arraut, Europhys.Lett. 109 (2015) 0002.
\bibitem{HawkingdRGT2} I. Arraut, arXiv:1503.02150 [gr-qc].
\bibitem{Kodama}  H. Kodama and I. Arraut, Prog. Theor. Exp. Phys. 023E02, (2014).
\bibitem{Higgs} I. Arraut, Europhys.Lett. {\bf111} (2015) 61001. 
\bibitem{Higgs2} I. Arraut, arXiv:1505.06215 [gr-qc]. 
\bibitem{My paper1} I. Arraut, Int.J.Mod.Phys. D24 (2015) 03, 1550022. 
\bibitem{My paper12} I. Arraut, Universe 3 (2017) 45.
\bibitem{My paper1la} I. Arraut, arXiv:1504.00467 [gr-qc].
\bibitem{Komar} I. Arraut, Phys.Rev. D90 (2014) 124082. 
\bibitem{V1} C. Mazuet, M. S. Volkov, Phys.Lett. {\bf B751} (2015) 19-24.
\bibitem{V2} M. S. Volkov, Lect.Notes Phys. {\bf892} (2015) 161-180.
\bibitem{V3} M. S. Volkov, Phys.Rev. D{\bf90} (2014) 12, 124090.
\bibitem{V4} M. S. Volkov, Phys.Rev. D{\bf90} (2014) 2, 024028. 
\bibitem{Publishedone} I. Arraut and K. Chelabi, EPL {\bf115} (2016) 31001.
\bibitem{Rubakov}  V. A. Rubakov and P. G. Tinyakov, Phys. Usp. \textbf{2008}, 51, 759--792.
\bibitem{Rubakov2} V. Rubakov, arXiv:hep-th/0407104.
\bibitem{Nima}  N. Arkani-Hamed, H.-C. Cheng, M. A. Luty and S. Mukohyama, JHEP 05, 074 (2004).
\end{thebibliography}
\end{document}